\documentclass[]{aastex7}
\usepackage{booktabs} \usepackage{siunitx} \usepackage{pgfplotstable} \usepackage{array}
\usepackage{longtable}
\usepackage{subfig}

\shorttitle{Stellar Paternity Tests}
\shortauthors{Schweers \& McSwain}
\graphicspath{{./}{figures/}}

\begin{document}

\title{Stellar Paternity Tests: Matching High-Latitude B Stars to the Open Clusters of their Birth}

\author[0009-0009-3021-6383]{Brandon Schweers}
\affiliation{Lehigh University 
Department of Physics, 16 Memorial Dr E, Bethlehem, PA 18015, USA}
\email{brandon.schweers240@gmail.com}

\author[0000-0002-4775-2803]{M.\ Virginia McSwain}
\affiliation{Lehigh University 
Department of Physics, 16 Memorial Dr E, Bethlehem, PA 18015, USA}
\email{mcswain@lehigh.edu}

\begin{abstract}

OB stars generally form in open clusters within the Milky Way's thin disk, so when they are found at high Galactic latitudes, it is thought that they were ejected from their birth clusters during the past few tens of millions of years. Using Gaia Data Release 3 (hereafter DR3) data, we traced the kinematic trajectories of 39 high-latitude B-type stars and 447 Galactic open clusters with high-quality astrometry to search for moments of past intersection.  In cases where we found matching trajectories, we also considered the clusters' HR diagrams to confirm parent-orphan pairs have matching ages. Further analysis of the clusters' core environments allowed us to determine a probable ejection mechanism. Through these paternity tests, we have identified possible origins for five of these orphaned B-type stars.  Here we present the likely travel times, ejection velocities, and a discussion of the runaway mechanism for each case.  We also identify one star whose trajectory did not bring it near the disk during the time period of our analysis, and we discuss its possible origins as well. 

\end{abstract}


\section{Introduction} 

About 25\% of Milky Way O stars are runaway stars with high peculiar space velocities and/or large distances from open clusters in the Galactic plane \citep{2023A&A...679A.109C, 1961BAN....15..265B}.  The frequency of runaways drops significantly among later spectral types, with B-type runaways occurring in only about 5\% of cases.  While B-type runaways occur less frequently, their longer lifetimes and more abundant numbers imply that they can still be numerous in the Milky Way.  

Most runaways were likely ejected from the open clusters of their birth by gravitational interactions in the dense cluster environment (the dynamical ejection scenario; DES).  Other runaway stars have likely acquired their high space velocities due to the conservation of momentum during the supernova of their close binary companion (the binary supernova scenario; BSS).
The DES dominates over BSS by a factor of 2-3 in the SMC \citep{2020ApJ...903...43D}, and \citet{2023A&A...679A.109C} argue that similar branching ratios probably apply in the Milky Way as well.

DES simulations consistently find that O-type stars are significantly overrepresented among ejected runaways relative to the Salpeter mass distribution, leaving the remaining cluster with a ``top-light'' distribution and steeper Salpeter slope (\citealt{2011Sci...334.1380F}; \citealt{2012ApJ...751..133P}; \citealt{2016A&A...590A.107O}).
\citet{2016A&A...590A.107O} showed that the high frequency of O-type ejections occurs because these massive stars sink into the cluster core within the first few Myr, increasing the cluster's core density and ejecting these massive stars with high efficiency.  Less massive B- and A-type stars are more likely to be ejected farther from the cluster center, where they are more abundant and the escape velocity is lower.  Ejection velocities up to 40 km~s$^{-1}$ are most typical for these stars since many trickle out as the cluster evaporates, although a few stars under 5 M$_\odot$ could exceed 200 or even 300 km~s$^{-1}$ if a massive star exchanges energy with a lower mass binary system.

For the BSS scenario, \citet{2011MNRAS.414.3501E} predicted the distribution of runaway velocities for both single stars and binaries that experience kick velocities in supernovae of various types.
They found that a type IIP supernova in a binary is the most likely type to involve a relatively low mass B-type secondary star with $M = 10.5 \pm 3.2\ \: M_\odot$.  The remaining secondary becomes unbound from the compact remnant and acquires a runaway velocity of $17.5 \pm 18.7$ km~s$^{-1}$ in this scenario.

Efforts to trace runaway stars back in time to identify their open cluster of origin have been limited to a few rare cases of nearby systems.   
\citet{1954ApJ...119..625B} were the first to identify a pair of stars, $\mu$ Columbae and AE Auriga, that were dynamically ejected from the Orion Nebula with approximately equal and opposite trajectories about 2.7 million years ago.  Improved astrometry with the Hipparcos mission allowed \citet{2001A&A...365...49H} to trace 22 runaway stars and 5 pulsars in the solar neighborhood back to their parent open clusters or associations.  And \citet{2005ApJ...621..978B} claimed that the stars HD 14633 and HD 15137 were ejected from NGC 654 in the Cas OB8 association, although their times of flight are paradoxically longer than the cluster's age.

Although high-quality distances and proper motions are now available in far greater numbers thanks to the Gaia mission, expanding the population of traceable runaways and parent clusters remains quite limited since radial velocities, $v_r$, are rarely available for hot stars.  Large spectroscopic surveys such as RAVE generally do not observe a large number of OBA stars since they avoid low Galactic latitudes to minimize source confusion in the fiber-fed spectrographs \citep{2020AJ....160...82S}.  The LAMOST survey did a better job of including OB stars in the plane, but their catalog of $\sim 16,000$ hot stars is highly concentrated in the outer Galactic disk \citep{2019ApJS..241...32L}.  
\citet{2023A&A...674A...7B} were able to extend high quality $v_r$ measurements in DR3 \citep{GaiaMission_2016A&A...595A...1G} to about a half million B- and A-type stars with $T_{\rm eff} < 14,500$ K and grvs\_mag $\leq 12$ mag in the Galactic plane, which greatly improves the 3-dimensional velocities for young open clusters.

Historically, another limiting factor in tracing the past trajectories has been the accuracy of Milky Way potential models.  \citet{2011MNRAS.411.2596S}, hereafter SN11, describe the complications of using fully continuous and analytical, but pre-Hipparcos, models of the Galactic disk, bulge, and halo \citep{1991RMxAA..22..255A}.  
SN11 used these Galactic potential models to analyze a modest sample of 96 B-type stars located at high Galactic latitudes and available high-quality $v_r$ measurements.  Through analysis of their spectral line features, SN11 conclude that these 96 stars are bona fide main-sequence stars.  They were able to trace all but three of them to an origin in the Galactic disk, estimating their ejection velocities and travel times, but the 3-D velocity measurements and Galactic potential models were not precise enough at the time to pinpoint a more specific origin.  

In this paper, we apply modern DR3 measurements and updated potential models to the SN11 sample of high-latitude B-type stars to attempt to trace them back to open clusters in the thin disk.  We describe the B star targets and open cluster sample in Section 2.  In Section 3, we model the kinematic trajectories of both groups through a modern Milky Way potential model to identify times of past intersections.  We further investigate the open cluster core environments and ages, including the possibility of blue stragglers, in Section 4.  We conclude in Section 5 with a discussion of our candidate matches and the characteristics of their ejections.  Overall, we find that the error bars on the 6-dimensional position and kinematic data are still too large to confidently pinpoint their origins, but we demonstrate a promising technique that will enable better matches in the future.

\section{Our Target Sample}

\subsection{High Latitude B Stars}

We started with the catalog of 96 high-latitude B-type stars, all of which were verified to be main-sequence stars, from SN11.  We used the DR3 to obtain astrometric and kinematic data for all targets. Many of these stars do not have $v_r$ measurements in the DR3, so for consistency, we used the $v_r$ measurements from SN11 in all cases. 

Most of our targets have distance measurements, $r_{GSP}$, in DR3 that are derived using the GSP-Phot algorithm.  Since these target stars lie well outside of the thin disk, their reddening is generally small and $r_{GSP}$ values are therefore more reliable.  However, for 48 targets, the parallax quality $\pi/\sigma_\pi < 10$, so both $\pi$ and GSP-Phot distances are unreliable.  We removed these stars from our target list. We also chose to exclude any star missing a $r_{GSP}$ value in DR3, which left us with 39 stars. We also obtained the astrometric quantities right ascension ($\alpha$), declination ($\delta$), and proper motions ($\mu \alpha \cos (\delta)$ and $\mu_\delta$); apparent magnitudes in the $G$, $BP$, and $RP$ bands; extinction ($A_G$); and reddening ($E(B-R)$) for these stars from DR3, along with their associated errors.

\subsection{Open Clusters}

We sourced the catalog of 1229 open clusters with over 400,000 possible member stars by \citet{OpenClusterMembersCatalog_2018A&A...618A..93C}, and we cross-referenced DR3 to obtain the same data for our cluster members.  We restricted our sample to only those stars with a membership probability of 80\% or higher.  This threshold minimizes the scatter in the clusters' properties while maintaining a large sample of cluster members.

For each open cluster, we determined its center by fitting Gaussian curves to the histograms of $\alpha$ and $\delta$ to calculate their mean values and $1\sigma$ error bars.  Clusters with fewer than 50 members could not be accurately fit with a Gaussian, so we excluded these clusters from our study.

\begin{figure}
    \centering
    \includegraphics[width=\linewidth]{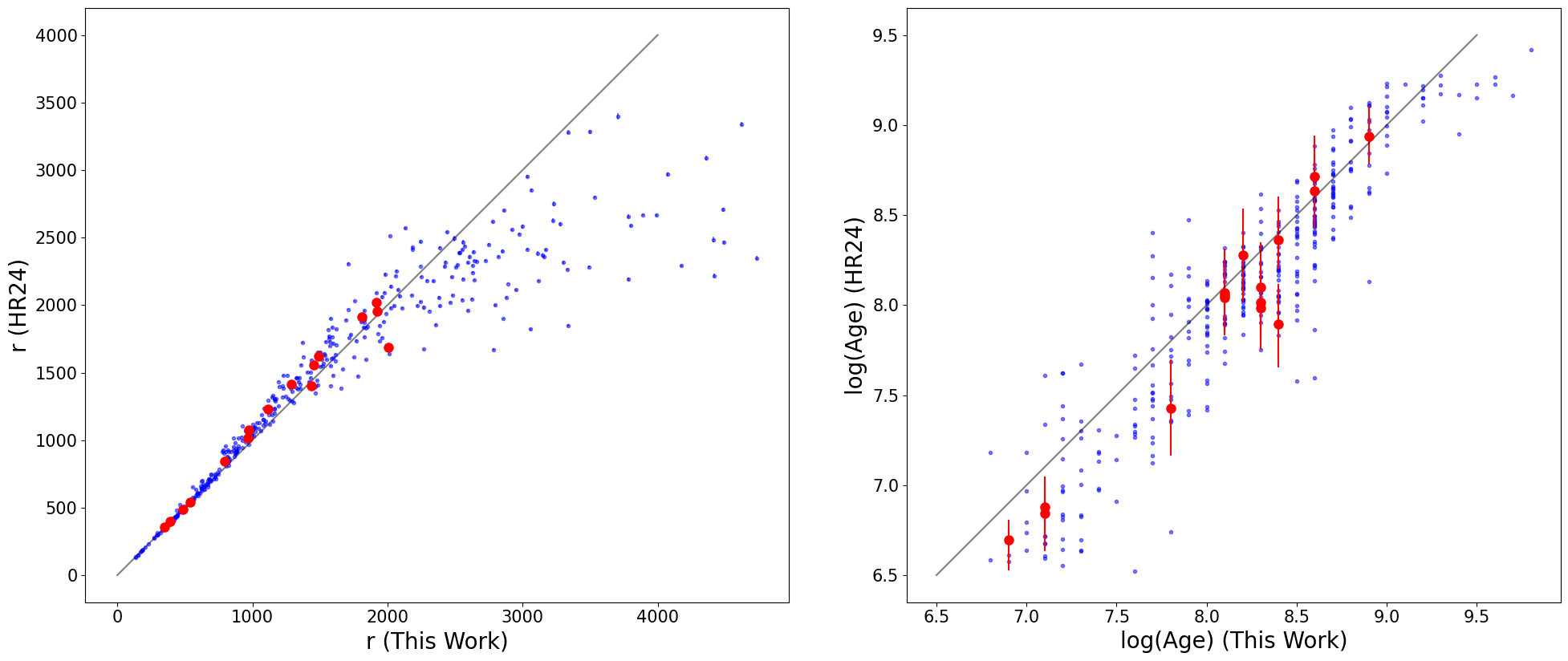}
    \caption{Our values for cluster $r$ and $\log$(Age) vs. the values found by HR24. The red points represent the clusters listed in our final results, while the blue points represent the remaining sample. In the left panel, the errorbars are all consistent with or smaller than the blue point sizes. In the right panel, the errorbars are shown for the red points only, but are typical in size for the larger sample. The gray line illustrates the identity function for reference.
    \label{fig:HR24_compare} 
    }
\end{figure}

Many of the cluster members did not have an available $r_{GSP}$ in DR3, so we require at least 60\% of total members to have this measurement for the cluster to be included in our study. With these members, we calculated a weighted mean distance ($r$) and standard deviation ($\Delta r$). For those 415 clusters that also had distances presented by \citet{2024A&A...686A..42H} (hereafter HR24), we compare our results in Figure \ref{fig:HR24_compare}. Our $r$, derived from $r_{GSP}$, shows excellent agreement for clusters with $r\lesssim2000$ pc but may be overestimated for the most distant clusters.

After determining the cluster distances, we turned to their radial velocities. Only about 12\% of the known cluster members had $v_r$ in DR3, so using $v_r$ from DR3 would require us to exclude a large portion of potential parent clusters. To preclude the need for $v_r$ measurements for every cluster, we assume that the clusters are confined to circular orbits in the Galactic disk with a flat rotation curve of circular speed $\Theta_0 = 236\pm11$ km~s$^{-1}$ \citep{v0_2009ApJ...704.1704B} and distance from the local standard of rest to the Galactic center $R_0 = 8.178\pm0.013$ kpc \citep{R0_2019A&A...625L..10G}.

Next, we determined the density profile of each cluster.  We measured the tidal radius ($R$) using the extrema of $\alpha$ and $\delta$ for the cluster members.
To calculate a core density, $\rho_c$, we followed the procedure used in \citet{2022A&A...659A..59T}. For each cluster member, we calculated its radial distance in the plane of the sky relative to the cluster center.  These radial distances were then sorted into bins of fixed widths. For the innermost $10$ pc, we used bins $0.5$ pc wide. For the next $20$ pc we used a bin width of $2$ pc, and for the outermost regions, we used bin widths of $10$ pc. This gives us $36$ bins spanning a $90$ pc distance and with greater resolution closer to the clusters' centers. If any of the bins in the inner $10$ pc of the cluster were empty, they were combined with their adjacent neighbor.

We calculated the surface density of each annulus and fit a King function \citep{1962AJ.....67..471K} to the density profile, using SciPy's ODR method \citep{SciPy}. This gives us values for the core radius ($R_c$), error ($\Delta R_c$), and the density ($\rho_c$) at that radius. For the error in this measurement ($\Delta\rho_c$), we used the square root of the density in accordance with a Poisson error\citep{poisson_error}.

We also measured the ages of these clusters to within $\pm 0.1$ dex using a grid of isochrones sourced from the online CMD 3.8 interface\footnote{The CMD 3.8 input form is available at https://stev.oapd.inaf.it/cgi-bin/cmd.}.  Our calculated isochrones used PARSEC version 1.2S \citep{Bressan_2012,Chen_2015} with solar metallicity and the Kroupa initial mass function \citep{kroupa_2001,kroupa_2002}.  We also used the Gaia EDR3 photometric system for ease of comparison to the observations with the YBC+new Vega setting (\citealt{Bohlin_2020}, \citealt{chen_2019}, and references therein).  Since we were largely concerned with identifying the main sequence turnoff, we used default parameters on CMD 3.8 for other evolved stars.  We generated a grid of isochrones between log age of 6 to 10 with a step size of 0.1 and compared them to each cluster's dereddened HR diagram.  Not all cluster members had a DR3 reddening value available so in cases where it was missing, we used the cluster's mean value.  We were unable to measure reliable ages for a few clusters that were either too sparse or showed evidence of clumpy reddening such that some of the brightest main sequence members could not be dereddened well from the cluster mean.  Although B-type stars should not live longer than 1 Gyr, we chose this range of ages to accurately identify an isochrone for all the clusters that remained in our sample.  There is the possibility that an older cluster could harbor blue stragglers, so we did not rule out origins in older clusters. We find good agreement between our age results and the results from HR24 as illustrated in the right panel of Figure \ref{fig:HR24_compare}.  

After removing the clusters that did not pass the criteria above, we were left with a sample of 447 remaining open clusters. Our calculated values for $\alpha$, $\delta$, $r$, $R$, $R_c$, $\rho_c$, $\log$(Age) and their respective errors are summarized in Table \ref{tab:clusters}.

\begin{deluxetable*}{lccccccccccccccccccc}

\tabletypesize{\tiny}

\tablecaption{Calculated properties of each cluster \label{tab:clusters}}

\tablenum{1}

\tablehead{\colhead{Cluster} & \colhead{$\alpha$} & \colhead{$\Delta\alpha$} & \colhead{$\delta$} & \colhead{$\Delta\delta$} & \colhead{$r$} & \colhead{$\Delta r$} & \colhead{$R$} & \colhead{$\Delta R$} & \colhead{$R_c$} & \colhead{$\Delta R_c$} & \colhead{$\rho_c$} & \colhead{$\Delta \rho_c$} & \colhead{$\log($Age$)$} \\ 
\colhead{} & \colhead{($^\circ$)} & \colhead{($^\circ$)} & \colhead{($^\circ$)} & \colhead{($^\circ$)} & \colhead{(pc)} & \colhead{(pc)} & \colhead{(pc)} & \colhead{(pc)} & \colhead{(pc)} & \colhead{(pc)} & \colhead{(stars pc$^{-2}$)} & \colhead{(stars pc$^{-2}$)} & \colhead{(Myr)}} 

\startdata
ASCC 10 & 51.85 & 0.24 & 34.96 & 0.27 & 639 & 68 & 29.44 & 2.87 & 1.27 & 1.13 & 0.8 & 0.05 & 8.6 \\
ASCC 101 & 288.33 & 0.24 & 36.35 & 0.26 & 390 & 18 & 13.44 & 2.09 & 3.09 & 1.76 & 0.9 & 0.1 & 8.2 \\
ASCC 105 & 295.54 & 0.5 & 27.41 & 0.58 & 541 & 49 & 34.45 & 1.49 & 1.15 & 1.07 & 1.4 & 0.43 & 8.5 \\
ASCC 108 & 298.34 & 0.33 & 39.4 & 0.45 & 1082 & 115 & 50.7 & 1.47 & 0.31 & 0.56 & 6 & 1.2 & 8.4 \\
ASCC 11 & 53.05 & 0.26 & 44.87 & 0.17 & 789 & 87 & 33.63 & 0.69 & 5.07 & 2.25 & 1.3 & 0.1 & 8.4 \\
ASCC 110 & 300.74 & 0.08 & 33.49 & 0.14 & 1831 & 305 & 39.49 & 7.37 & 0.47 & 0.68 & 2.6 & 0.42 & 8.8 \\
ASCC 111 & 302.94 & 0.32 & 37.52 & 0.3 & 798 & 73 & 46.09 & 3.09 & 1.86 & 1.36 & 1.2 & 0.15 & 8.4 \\
\ldots & 
\ldots & 
\ldots & 
\ldots & 
\ldots & 
\ldots & 
\ldots & 
\ldots & 
\ldots & 
\ldots & 
\ldots & 
\ldots & 
\ldots & 
\ldots \\
Trumpler 30 & 269.17 & 0.09 & -35.3 & 0.08 & 1292 & 313 & 7.63 & 0.32 & 1.92 & 1.39 & 2.3 & 0.14 & 8.6 \\
Trumpler 32 & 274.3 & 0.04 & -13.35 & 0.04 & 1752 & 358 & 7.61 & 0.74 & 4.57 & 2.14 & 1 & 0.03 & 8.6 \\
Trumpler 7 & 111.85 & 0.05 & -23.94 & 0.05 & 1488 & 328 & 6.89 & 0.15 & 5.3 & 2.3 & 1 & 0.07 & 8.1 \\
Trumpler 9 & 118.91 & 0.03 & -25.89 & 0.03 & 2560 & 549 & 11.23 & 0.24 & 3.84 & 1.96 & 1.4 & 0.22 & 7.7 \\
vdBergh 80 & 97.77 & 0.15 & -9.62 & 0.07 & 799 & 181 & 8.35 & 0.13 & 2.75 & 1.66 & 1.1 & 0.46 & nan \\
vdBergh 92 & 106.03 & 0.07 & -11.47 & 0.12 & 924 & 225 & 8.5 & 0.07 & 2.54 & 1.59 & 2.4 & 0.47 & 7.2 \\
\enddata

\tablecomments{Table \ref{tab:clusters} is published in its entirety in the machine-readable format. A portion is shown here for guidance regarding its form and content.}

\end{deluxetable*}

\section{Matching the Kinematic Trajectories}

Using the galpy Python package with the 2014 Milky Way potential \citep{2015ApJS..216...29B}, we modeled the gravitational potential of the Milky Way and calculated the past trajectories of our orphan stars. We chose a timescale of 30 Myr (with possible extension to 50 Myr; see below) to minimize error in the trajectories while also covering a large fraction of these B-type stars' main sequence lifetimes ($\sim 30-100$ Myr).

In choosing such a lengthy timescale, we introduce the possibility that strong and weak encounters with other stars or molecular clouds may alter our targets' trajectories and cause faulty results. Open clusters are much more massive than any individual star they may encounter in their orbits. As such, open clusters are less likely to have their trajectories altered.  The high-latitude stars are not as resistant to perturbations as the open clusters, so we must ensure that such encounters are unlikely to have altered the results of our study. 

The high-latitude stars are currently found at large distances above the Galactic plane, so they must have traveled roughly perpendicularly out from the disk.  The thin disk has a scale height $\sim 300$ pc, while the less dense thick disk extends to $\sim 1000$ pc \citep{BinneyAndMerrifield}.  Within the thin disk, an ejected star will likely experience many weak gravitational perturbations from other stars until it reaches the very low-density region beyond the scale height.  These encounters would preferentially nudge the star along the radial and azimuthal axes of the Galaxy, perpendicular to the star's direction of travel, in such a way that they average out to zero.  The mean-square change in the star's velocity during this single disk crossing can be estimated as
\begin{equation}
    \langle\Delta V_\perp^2\rangle = \frac{8\pi G^2m^2nt}{V} \ln \left( \frac{b_{max}}{b_{min}} \right)
\end{equation}
\citep{BinneyAndTremaine} where $t$ is the crossing time, $b$ is the interaction distance, $V$ denotes the velocity of the target star relative to nearby stars, $m$ is the typical mass of a nearby star, $n$ is the stellar density of the region, and $G$ is the gravitational constant. We chose a range of interaction distances $.01 \le b \le 10\ pc$.  A star traveling at $V = 30\ km\ s^{-1}$ will take approximately $10\ Myr$ to cross the $300\ pc$ thin disk.  We conservatively set $n \sim 1\ star~pc^3$ and $m = 5\ M_\odot$, and find $\langle\Delta V_\perp^2\rangle \sim .026$ km$^2$ s$^{-2}$.  Any change in the ejected star's trajectory is thus small enough to be well within the measured error bars, so we expect the cumulative effect of weak encounters in the thin disk to be negligible.
As the star crosses into the thick disk, its lower density results in even fewer perturbations on the stellar trajectory.

\begin{figure}
    \centering
    \includegraphics[width=0.75\linewidth]{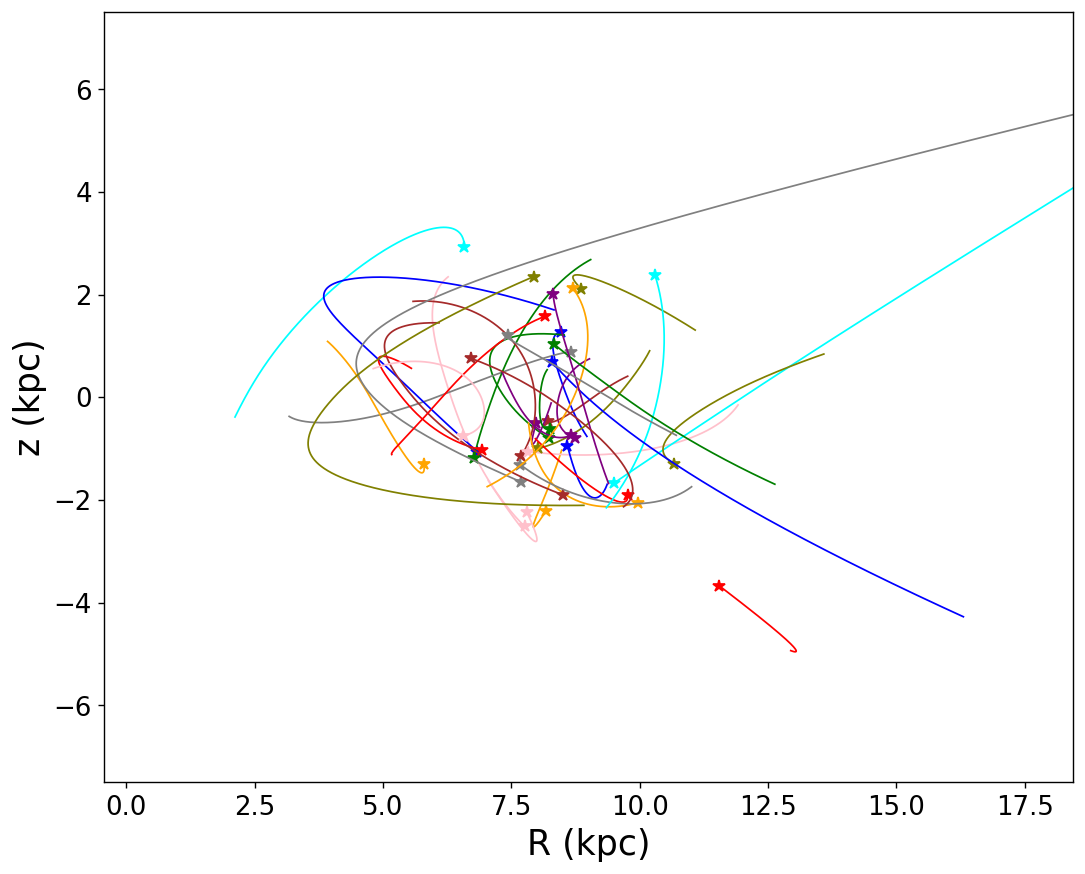}
    \caption{The edge-on view of the trajectories for our 39 remaining target stars. Their current positions are marked by the star symbols, and the trajectories are extended to 50 Myr in the past for illustrative purposes.
    \label{fig:all_star_trajectories} 
    }
\end{figure}

For each target star and open cluster, we generated its default trajectory as well as 99 randomized trajectories using a normal distribution for each of its position and velocity parameters and our measured $\pm 1\sigma$ error bars.  This allowed us to fully account for the uncertainties in their possible trajectories.  Not all of these randomized trajectories for the orphan stars crossed the disk plane within 30 Myr, so in cases where some of the random trajectories did not, we extended the integration to longer times to allow all of the random trials to encounter the disk plane.  Figure \ref{fig:all_star_trajectories} illustrates a cross-section of the Galactic disk and the extended trajectories of our target stars.

\begin{figure}
    \centering
    \includegraphics[width=0.75\linewidth]{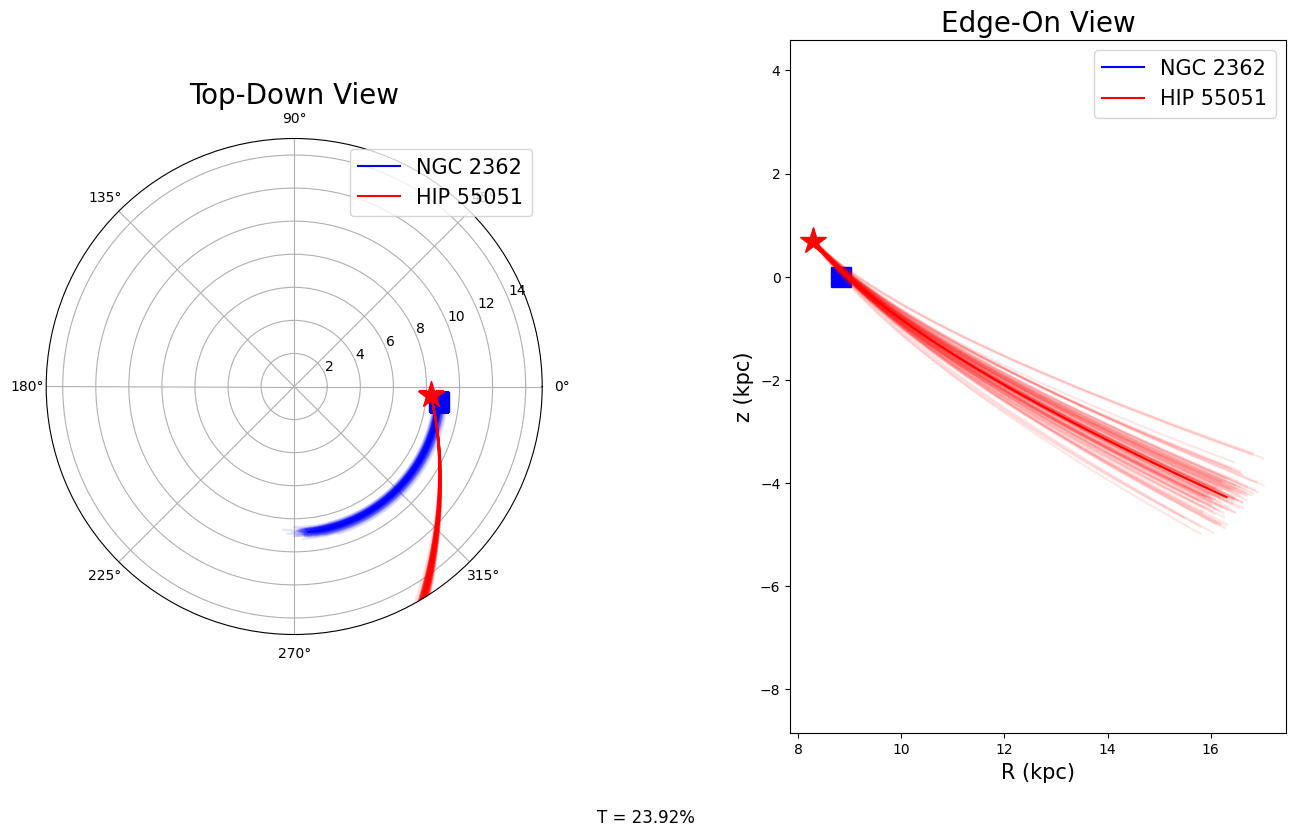}
    \caption{The trajectories of the star HIP 55051 (red) and the cluster NGC 2362 (blue) in Galactocentric coordinates are shown as an example of a candidate match with $T = 23.92$\%.  The star and square markers indicate their current positions. While these trajectories extend to 50 Myr in the past for illustrative purposes, the intersection points for this pair occur $\sim5.7\pm.4$ Myr ago.
    \label{fig:trajectory_good}
    }
\end{figure}

With our calculated trajectories, we searched for all times of past intersection within a threshold distance of 100 pc between an orphan star and a potential parent cluster. We calculated a match probability ($T$) for each pair based on the number of observed intersections in relation to a possible 10,000 intersections. Any match where $T \ge 10$\% was saved as a potential candidate. Figure \ref{fig:trajectory_good} shows an example of a star-cluster pair with a high $T$ value.

For each candidate match, we also calculated the ejection velocity of the runaway star by computing its velocity relative to the potential birth cluster at the time of intersection.  For the full range of possible trajectory matches, these resulted in a spread of $v_{eject}$ and $t_{eject}$ values for each cluster-runaway match. We report the mean and standard deviation of these spreads in Table \ref{tab:results} as $v_{eject}$ and $\Delta v_{eject}$, and $t_{eject}$ and $\Delta t_{eject}$, respectively.

\section{The Open Cluster Environments}

Our analysis of the kinematics of high-latitude stars found that many of these ejected stars have multiple candidate parent clusters, so many of the trajectory matches are likely to be spurious.  To rule out false matches, we analyzed the combined color-magnitude diagrams of each open cluster with its potential runaway star. 

For each orphan-parent pair that we saved, we plotted the orphan star and the cluster members on an HR diagram, along with the cluster's isochrone. We categorized the pairs into three groups (yes, no, and maybe), and the results of this categorization can be found under the Age Agreement column in Table \ref{tab:results}. 

\begin{figure}
    \centering
    \includegraphics[width=\linewidth]{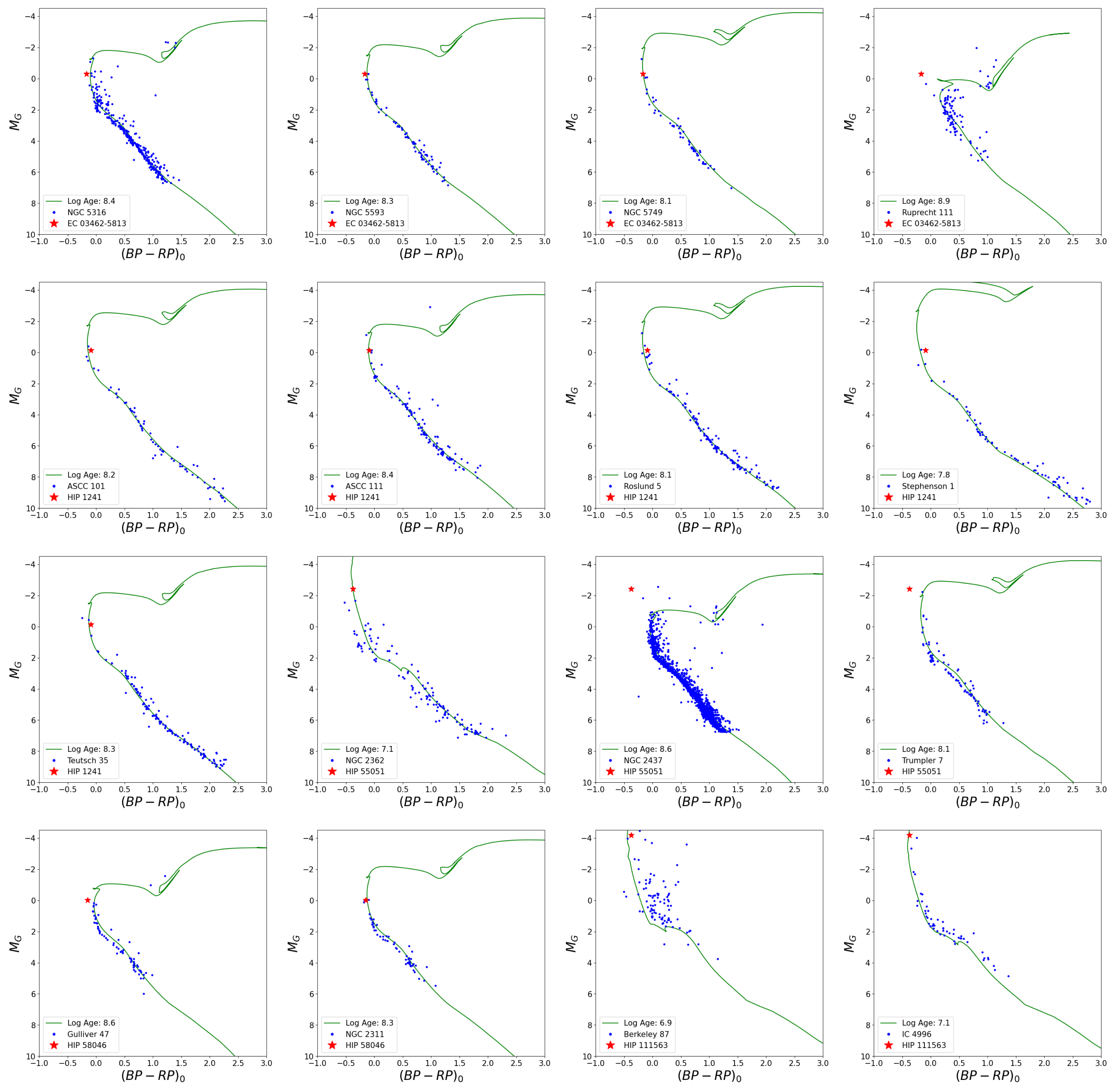}
    \caption{The HR diagrams for all candidate matches. Three of these cases were identified as potential blue stragglers: EC 03462-5813 with Ruprecht 111, HIP 55051 with NGC 2437, and HIP 55051 with Trumpler 7. All other orphan stars had ages consistent with their parent clusters.
    \label{fig:hr_spread}
    }
\end{figure}

We assigned a `y' to pairs that showed strong agreement in their ages.  Matches marked with an `m' included stars that may be blue stragglers, which we required to be less than 2 magnitudes brighter than the main sequence turnoff \citep{2009Natur.462.1032M}. The third group contained the pairs that showed poor agreement in their ages and was assigned an `n' value.  The panels in Figure \ref{fig:hr_spread} illustrate the `y' and `m' matches.

To further refine our matching results, we considered the clusters' core densities, $\rho_c$, and core radii, $R_c$. Since a cluster with a higher $\rho_c$ or smaller $R_c$ will have more strong gravitational interactions between its core members, it will be more likely to eject a star via dynamical ejection than a sparse cluster. For these parameters, we set a threshold of $\rho_c \geq 5$ stars~pc$^{-2}$ or $R_c \leq 1.5$ pc to categorize our proposed DES candidates. 

Ejection via a supernova could take place in either a rich or sparse environment, so we classified matches that had good age agreement but did not meet either of the core thresholds as consistent with a BSS. Since a BSS requires a supernova explosion from a B2 or earlier spectral type, we also set an upper limit of $\log($Age$)=8.0$ for clusters considered as BSS origin sites. This allows sufficient time for a star ejected at the beginning of our 30 Myr timescale to have potentially come from a supernova explosion.

Blue straggler stars are thought to be rejuvenated by stellar mergers \citep{1976ApL....17...87H} or mass transfer between close binaries \citep{1964MNRAS.128..147M}.  Since blue stragglers are observed to be binaries at a higher rate than normal stars \citep{2009Natur.462.1032M}, and mass segregation is expected to cause close massive binaries to settle in the cluster core, both DES and BSS ejections are possible mechanisms for matches in the `m' category.

\clearpage
\clearpage

\section{Conclusions}
In this study, we have developed a sort of ``paternity test'' for determining a runaway star's cluster of origin. We analyzed the kinematic trajectories of the high-latitude B stars and open clusters to search for past intersections, and we compared their ages using an HR diagram. Further analysis of the clusters' core parameters provided a diagnostic to probe the ejection method. Of the 39 high-latitude B-type stars that we considered here, we found five unique stars that were strong candidates to match with a total of sixteen parent clusters. Table \ref{tab:results} summarizes our findings.  We also note one very high latitude star whose trajectory did not bring it near the disk in our analysis, and we discuss it further below.

\begin{longrotatetable}
\begin{deluxetable}{llcccccccccl}

\tabletypesize{\scriptsize}

\tablecaption{Ejection probabilities and characteristics \label{tab:results}}

\tablenum{2}

\tablehead{\colhead{Star} & \colhead{Cluster} & \colhead{$T$} & \colhead{$v_{eject}$} & \colhead{$\Delta v_{eject}$} & \colhead{$t_{eject}$} & \colhead{$\Delta t_{eject}$} & \colhead{Cluster $\log($Age$)$} & \colhead{Age} & \colhead{$\rho_c$} & \colhead{$R_c$} & \colhead{Match Comments} \\ 
\colhead{} & \colhead{} & \colhead{($\%$)} & \colhead{($km\ s^{-1}$)} & \colhead{($km\ s^{-1}$)} & \colhead{($Myr$)} & \colhead{($Myr$)} & \colhead{($Myr$)} & \colhead{Agreement} & \colhead{($stars\ pc^{-2}$)} & \colhead{($pc$)} & \colhead{} } 

\startdata
EC 03462-5813 & NGC 5316 & 11.09 & 90.6 & 7.6 & -13.3 & 1.2 & 8.4 & y & 4.66 & 2.2 & Consistent with DES \\
 & NGC 5593 & 13.52 & 94 & 8 & -11.8 & 1.1 & 8.3 & y & 3.47 & 1.1 & Consistent with DES \\
 & NGC 5749 & 14.61 & 92.9 & 8.8 & -12 & 1.2 & 8.1 & y & 3.2 & 0.8 & Consistent with DES \\
 & Ruprecht 111 & 13.57 & 90.3 & 8 & -12.9 & 1.2 & 8.9 & m & 5.17 & 1.1 & Blue straggler. Potential DES \\
HIP 1241 & Alessi 12 & 16 & 37.4 & 1 & -31.2 & 3 & 8.2 & y & 1.24 & 3.8 & Both DES and BSS unlikely \\
 & ASCC 101 & 17.22 & 37 & 0.8 & -30.2 & 2.7 & 8.2 & y & 3.09 & 0.9 & Consistent with DES \\
 & ASCC 111 & 18.08 & 38.8 & 1 & -31.6 & 3 & 8.4 & y & 1.86 & 1.2 & Consistent with DES \\
 & Roslund 5 & 20.83 & 37.5 & 0.9 & -30.7 & 2.8 & 8.1 & y & 2.98 & 1.5 & Consistent with DES \\
 & Stephenson 1 & 20.47 & 37.1 & 0.9 & -30.8 & 2.9 & 7.8 & y & 6.37 & 0.8 & Consistent with DES \\
 & Stock 1 & 13.6 & 37.4 & 1 & -31.7 & 3 & 8.5 & y & 1.71 & 2 & Both DES and BSS unlikely \\
 & Teutsch 35 & 20.59 & 37.3 & 0.9 & -30.9 & 2.8 & 8.3 & y & 3.79 & 1.2 & Consistent with DES \\
HIP 55051 & Haffner 5 & 33.11 & 157.7 & 5.3 & -6 & 0.4 & 9.5 & n & 11.49 & 1 &  \\
 & NGC 2354 & 40.5 & 158.5 & 5.4 & -5.6 & 0.3 & 9 & n & 3.04 & 2.5 &  \\
 & NGC 2362 & 23.92 & 158.6 & 5.4 & -5.7 & 0.4 & 7.1 & y & 10.78 & 0.8 & Consistent with DES \\
 & NGC 2428 & 27.37 & 158.8 & 5.2 & -6.1 & 0.4 & 8.7 & n & 3.08 & 1.5 &  \\
 & NGC 2432 & 24.33 & 157.7 & 5.5 & -6.2 & 0.4 & 8.8 & n & 12.49 & 0.8 &  \\
 & NGC 2437 & 10.46 & 159.6 & 5.3 & -6.2 & 0.4 & 8.6 & m & 13.13 & 2.4 & Blue straggler. Potential DES \\
 & NGC 2448 & 12.49 & 158 & 5.6 & -5.4 & 0.2 & 8.4 & m & 0.47 & 4.7 & Both DES and BSS unlikely \\
 & NGC 2482 & 18.75 & 156.7 & 5.4 & -5.6 & 0.3 & 8.7 & n & 1.21 & 2.6 &  \\
 & NGC 2539 & 25.49 & 158.5 & 5.2 & -5.9 & 0.4 & 8.8 & n & 8.11 & 1.5 &  \\
 & Trumpler 7 & 23.22 & 157.9 & 5.5 & -6 & 0.4 & 8.1 & m & 5.3 & 1 & Blue straggler. Potential DES \\
HIP 58046 & Gulliver 47 & 12.02 & 100.8 & 2.4 & -18.4 & 0.7 & 8.6 & y & 1.21 & 1.4 & Consistent with DES \\
 & NGC 2286 & 10.31 & 99.8 & 2 & -18 & 0.9 & 8.6 & m & 3.07 & 1.6 & Both DES and BSS unlikely \\
 & NGC 2311 & 13.88 & 98 & 2.1 & -18.4 & 0.8 & 8.3 & y & 4.06 & 0.9 & Consistent with DES \\
HIP 81153 & Gulliver 33 & 15.12 & 151.6 & 7 & -10.8 & 0.5 & 8.4 & n & 0.26 & 4.8 &  \\
 & NGC 7082 & 11.32 & 154.1 & 6.8 & -10.9 & 0.5 & 8.4 & n & 5.67 & 1.3 &  \\
 & Roslund 7 & 16.75 & 153.4 & 7.1 & -9.7 & 0.5 & 8.1 & m & 0.22 & 6.4 & Both DES and BSS unlikely \\
HIP 111563 & Berkeley 87 & 14.79 & 89 & 5.5 & -14.1 & 0.7 & 6.9 & y & 2.48 & 1.2 & Consistent with DES \\
 & IC 4996 & 19 & 88.7 & 5.2 & -14 & 0.8 & 7.1 & y & 9.08 & 0.6 & Consistent with DES \\
\enddata

\end{deluxetable}
\end{longrotatetable} 
\textit{EC 05438-4741} -- We identified one star, EC 05438-4741, whose trajectory does not appear to approach the thin disk even within an extended time frame of 50 Myr. Currently, this star lies $\sim 3.5$ kpc below the Galactic plane and about 12 kpc away from the Galactic center. Its current trajectory suggests that this star is moving towards the disk, rather than away from it, raising the question of whether EC 05438-4741 could have formed in situ in the halo \citep{2008ApJ...685L.125D}.  Its effective temperature $T_{\rm eff} = 13500$ K and surface gravity $\log g = 4.10$ (SN11; \citealt{1997MNRAS.290..422R}) indicate that it is a late B-type star that may spend as much as 100 Myr on the main sequence. After extending our trajectory calculations for this singular star to 100 Myr, we saw that this star appears to originate from the disk after all, removing the possibility of halo formation. We hesitate to claim any cluster of origin for such a long travel time.

\begin{figure}
    \centering
    \includegraphics[width=1\linewidth]{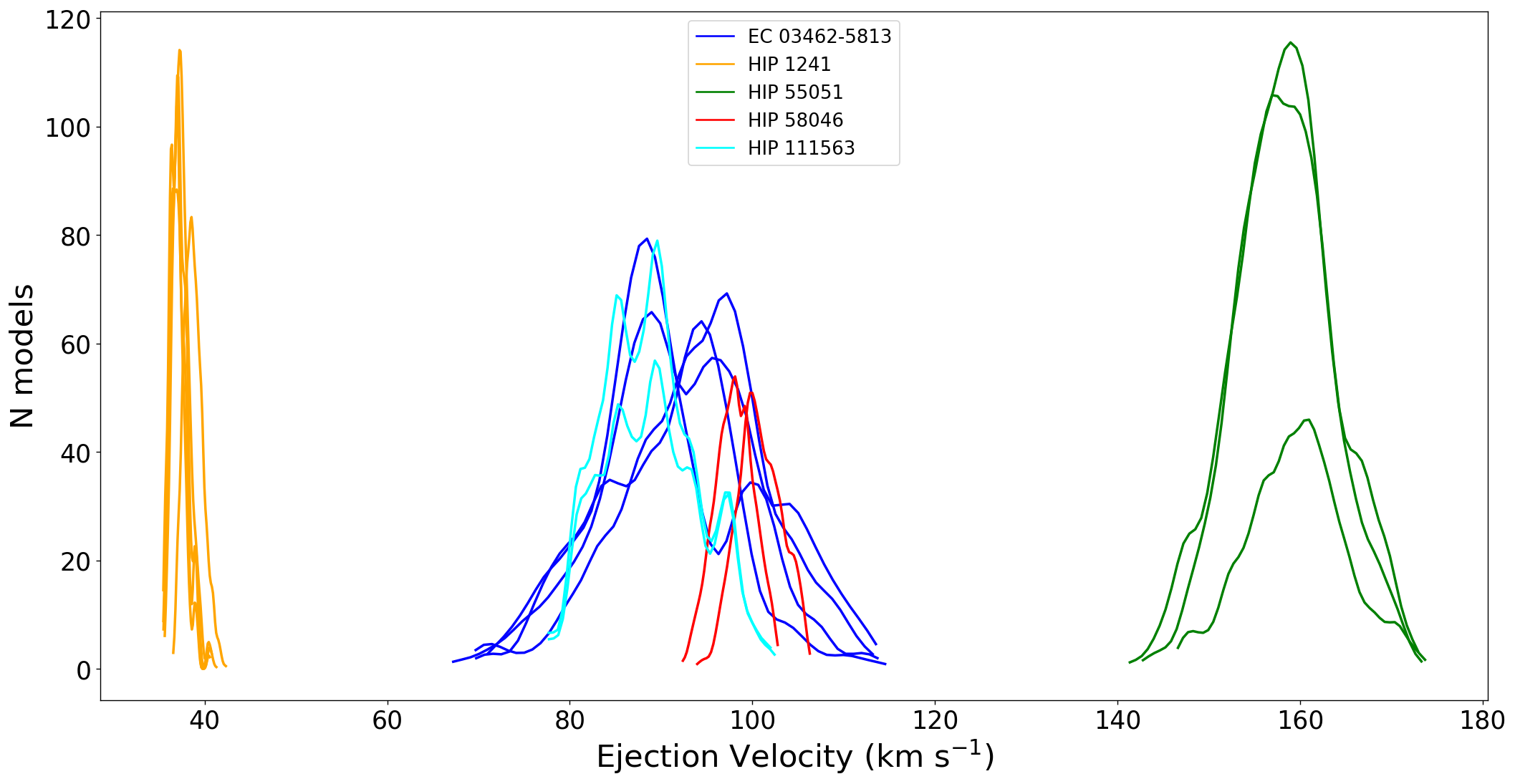}
    \caption{The smoothed histograms of $v_{eject}$ for our 16 candidate matches. Each curve represents the $v_{eject}$ for one orphan-parent pair.}
    \label{fig:v_smooth}
\end{figure}

\textit{EC 03462-5813} -- We found a likelihood of $T=11.09$\%, $T= 13.52$\%, and $T= 14.61$\% that EC 03462-5813 was ejected from NGC 5316, NGC 5593, or NGC 5749 respectively.  Due to the relatively small $R_c$ of NGC 5593 and NGC 5749, this ejection is more consistent with a DES interaction. We considered NGC 5316 to also be a DES candidate due to its $\rho_c=4.7\pm2$ placing it at our threshold for core density.  Although Ruprecht 111 is somewhat older, it is also a candidate parent ($T= 13.57$\%) of EC 03462-5813, making this a potential blue straggler DES match. EC 03462-5813's ejection would have occurred about $12.5 \pm 1.3$ Myr ago and with $v_{eject} \approx 92 \pm 8$ km s$^{-1}$.  The full range of possible $v_{eject}$ values is illustrated in Figure \ref{fig:v_smooth}. When EC 03462-5813 was ejected, the four potential parent clusters were separated by $\sim500$ pc.

\begin{figure}
    \centering
    \includegraphics[width=1\linewidth]{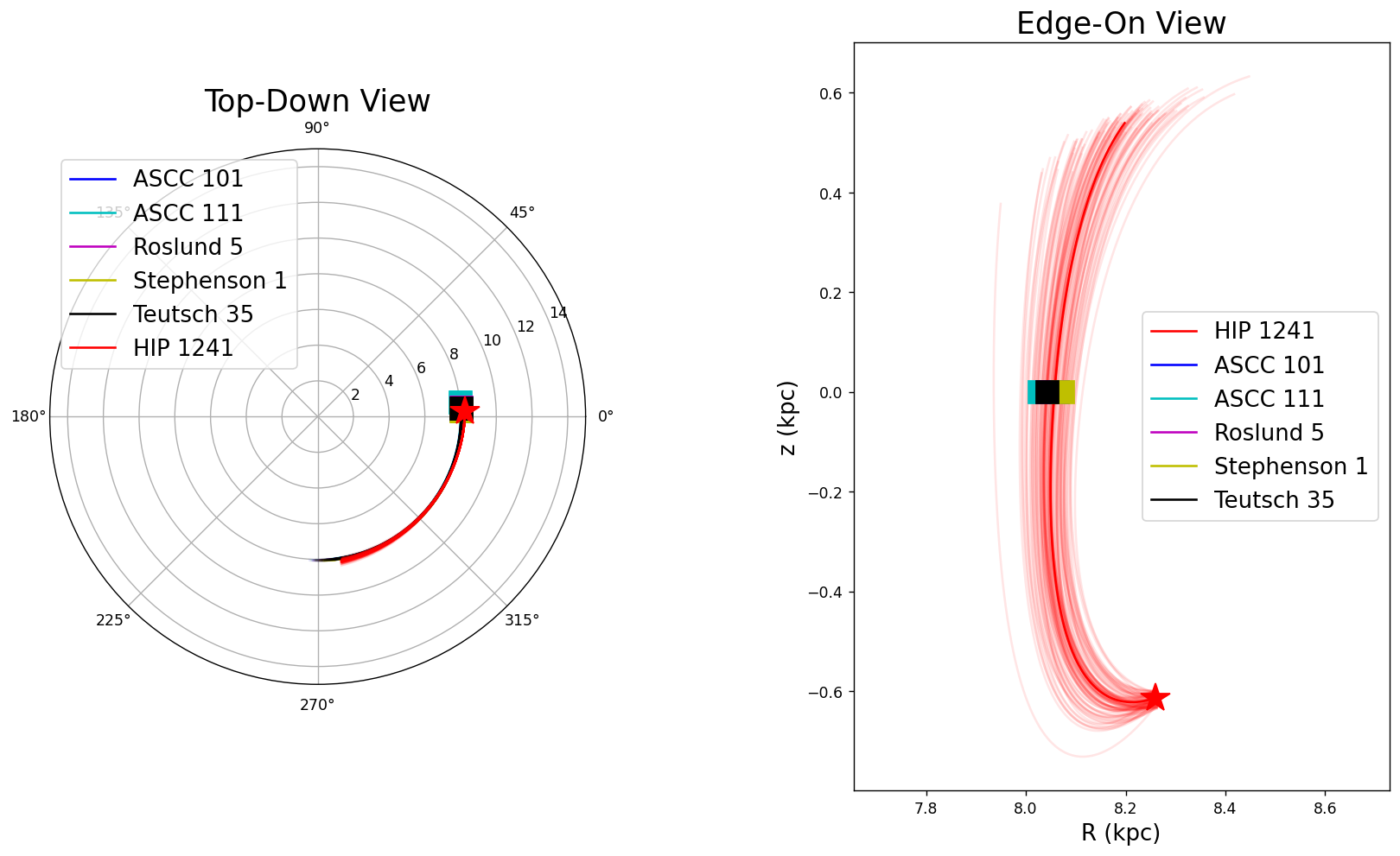}
    \caption{We present the trajectories of the star HIP 1241 with its five possible origin clusters in a similar format to Figure \ref{fig:trajectory_good}. With the assumption of circular motion, at the time of ejection, they were located within 500 pc of one another.}
    \label{fig:HIP_1241_clusters}
\end{figure}

\textit{HIP 1241} -- This star was found to have five candidate parent clusters. ASCC 101 ($T=17.22$\%), ASCC 111 ($T=18.08$\%), Roslund 5 ($T=20.83$\%), Stephenson 1 ($T=20.47$\%), and Teutsch 35 ($T=20.59$\%) all have small $R_c$, making them consistent with DES ejection sites. HIP 1241 also crossed paths with Alessi 12 and Stock 1 with good age agreement, but those clusters' low core densities and high $\log($Age$)$ made them ineligible for both DES and BSS consideration. Although we cannot pinpoint which cluster HIP 1241 came from, we can say with confidence it was ejected from the Galactic disk about $31 \pm 3$ Myr ago with $v_{eject} \approx 38 \pm 1$ km s$^{-1}$.  At the time of ejection, these clusters were within a $\sim500$ pc distance of each other, and this closeness is shown in Figure \ref{fig:HIP_1241_clusters}.

\textit{HIP 55051} -- We found ten clusters with good trajectory matches to HIP 55051, but only one cluster, NGC 2362 ($T=23.92$\%), had good age agreement. We categorized this as a DES candidate. Three other clusters were categorized as `m' in age agreement, and one those three is both too old and not dense enough to be considered for either BSS or DES. The remaining two, NGC 2437 ($T=10.46$\%) and Trumpler 7 ($T=23.22$\%) are consistent with blue straggler DES origin sites. This ejection likely occurred $6 \pm 0.5$ Myr ago with $v_{eject} \approx 160 \pm 5$ km s$^{-1}$, and at this time, the three clusters were within a $\sim400$ pc distance of each other. Having a $v_{eject} \approx 160$ km s$^{-1}$ makes this our highest velocity candidate.  \citet{2016A&A...590A.107O} found that ejection via dynamical interactions can reach and even exceed an ejection velocity of 100--200 km s$^{-1}$, which is consistent with our findings.

\textit{HIP 58046} -- We found HIP 58046 could have originated in Gulliver 47 ($T = 12.02$\%), NGC 2286 ($T=10.31$\%), or NGC 2311 ($T = 13.88$\%). Gulliver 47 and NGC 2311 have a small $R_c$ making them both potential sites for a DES ejection. NGC 2286 is both too sparse and too old to be a reasonable ejection site. At our estimated $t_{eject}$ of $18\pm1$ Myr ago, these clusters were separated by $\sim200$ pc. Whichever cluster HIP 58046 may have come from, its $v_{eject}$ was approximately $99\pm3$ km s$^{-1}$.

\textit{HIP 81153} -- Although HIP 81153 appears to have crossed paths with Gulliver 33, NGC 7082, and Roslund 7 with $T$ values of 15.12\%, 11.32\%, and 16.75\% respectively, we are unable to classify any of these pairings as a potential ejection site. A combination of old cluster age and lackluster core densities lead us to rule out these three clusters.  Despite not being able to point to a specific origin, we have determined that HIP 81153 likely originated in the Galactic disk and was ejected $\sim 10 \pm 1$ Myr ago. Since then, the star has been traveling away from the disk at a high speed of $\sim 150 \pm 10$ km s$^{-1}$.  This very high ejection speed prompted us to consider whether we might have accidentally excluded the parent cluster in our analysis.  We identified in SIMBAD several other sparse open clusters nearby, but none were included in the catalog of \cite{OpenClusterMembersCatalog_2018A&A...618A..93C}.  Since the cluster populations and  ages are unknown, we cannot conclusively match any with HIP 81153.

\textit{HIP 111563} -- Berkeley 87 ($T= 14.79$\%) has a small $R_c$ and IC 4996 ($T=19.00$\%) has a high $\rho_c$ in addition to a low $R_c$, making both of these clusters potential DES origin sites. An ejection from either of these clusters likely occurred about $14\pm1$ Myr ago and at $v_{eject} \sim 89\pm5$ km s$^{-1}$. At that time, these two clusters were only separated by $\sim100$ pc, making them our tightest origin site for an orphan star in this study. Our measured age of 12.6 Myr for IC 4996 is more consistent with this travel time than for the younger Berkeley 87. The messy HR diagram of Berkeley 87 supports a young cluster environment but obfuscates a precise age measurement. It is worth noting that both Berkeley 87 and IC 4996 appear to be members of OB associations in the Cygnus region. Berkeley 87 lies only 31 pc from Group D and 77 pc from Group F as identified in \citet{2021MNRAS.508.2370Q}, while IC 4996 is located $\approx75$ pc from both Groups D and F. It seems reasonable that HIP 111563 was ejected from one of these OB associations.

We plotted a smoothed histogram of $v_{eject}$ for all five stars and sixteen possible origins, shown in Figure \ref{fig:v_smooth}. We smoothed the data using a Gaussian kernel from the specutils Python library \citep{nicholas_earl_2024_14042033}.  The values of $v_{eject}$ are generally consistent with the expectations of the DES ejection.  

\begin{figure}
    \centering
    \includegraphics[width=1\linewidth]{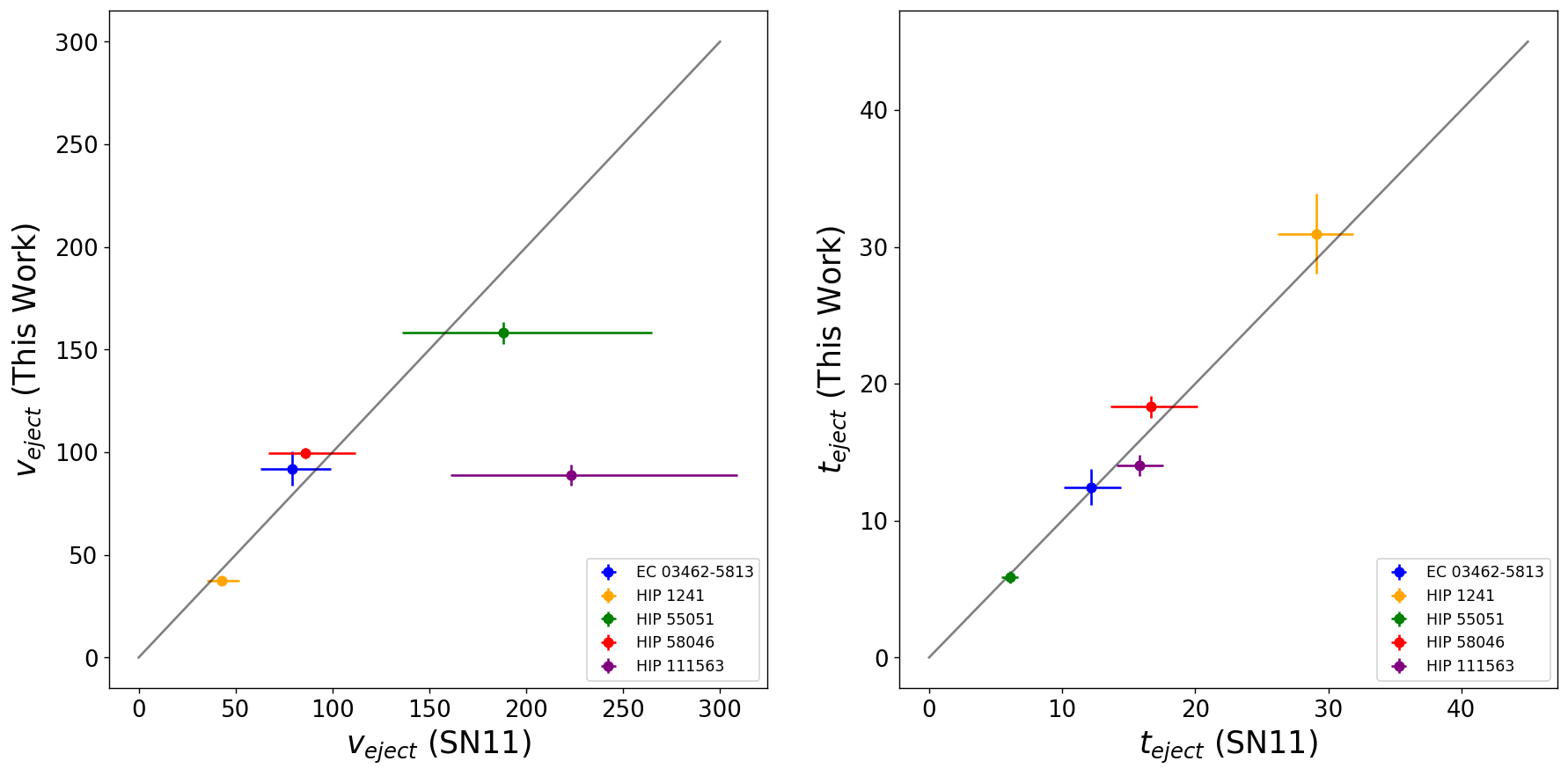}
    \caption{Our values for $v_{eject}$ and $t_{eject}$ vs.\ the values found by SN11. The gray line illustrates the identity function for reference.}
    \label{fig:silva_compare}
\end{figure}

Figure \ref{fig:silva_compare} shows a comparison of our weighted mean $v_{eject}$ and $t_{eject}$ with the values from SN11. We found our values to have good agreement with SN11 with the exception of the value of $v_{eject}$ for the star HIP 111563. SN11 calculated $v_{eject}$ to be more than double the value we calculated despite the agreement in $t_{eject}$ for this star. The most likely explanation for this discrepancy is the difference in $r$ used in both papers. SN11 calculated $r$ using the distance modulus of each star, while we are using the more recent $r_{GSP}$ from Gaia DR3.

Additionally, there are several very young, massive clusters (e.g. NGC 869, NGC 884, NGC 6242, NGC 7654, IC 1805, Trumpler 14-16) that we would expect to have ejected a star within the past 10 Myr that do not appear in our final results.  These clusters all passed our criteria for inclusion, but simply did not match this sample of runaways. We suspect this is because their recent ejections may not have had time to climb to the Galactic heights that our target sample has.  Further work is needed to explore more hot stars in the Galactic disk with high peculiar velocities to identify runaways from the youngest open clusters.

We recognize that the trajectory matching technique is still fraught with high error that limits our ability to identify parent clusters for our orphan stars. The bulk of this error comes from the $v_r$ measurements of the high latitude stars, not their Gaia astrometry, since the spectral line broadening of these hot stars precludes the ability to reduce the error bars in $v_r$. Furthermore, only $\sim 12$\% of the open cluster members have $v_r$ available in DR3, so the lack of reliable $v_r$ measurements for most open cluster members prohibits a robust trajectory calculation for the young clusters.  Overall, $v_r$ remains the largest limiting factor in our stellar paternity tests.

\begin{acknowledgments}
We are grateful for support from Thomas Cahill and Lehigh University.  We also thank Christopher Aviles Bramer for his contributions to a preliminary version of this study.  We appreciate the constructive comments from the anonymous referee that greatly improved this paper.  This research has made use of the VizieR catalogue access tool and the SIMBAD database, operated at CDS, Strasbourg, France.   This work has made use of data from the European Space Agency (ESA) mission
{\it Gaia} (\url{https://www.cosmos.esa.int/gaia}), processed by the {\it Gaia}
Data Processing and Analysis Consortium (DPAC,
\url{https://www.cosmos.esa.int/web/gaia/dpac/consortium}). Funding for the DPAC
has been provided by national institutions, in particular the institutions
participating in the {\it Gaia} Multilateral Agreement.
\end{acknowledgments}

\clearpage
\bibliography{new.ms}{}
\bibliographystyle{aasjournal}

\end{document}